\documentclass[10pt, conference]{IEEEtran}

\usepackage{color}
\usepackage[table]{xcolor}
\usepackage{graphicx}
\usepackage{amssymb}
\usepackage{amsmath}
\usepackage{graphicx}
\usepackage{latexsym}
\usepackage{epstopdf}
\usepackage{multicol}
\usepackage{caption}
\usepackage[vlined,ruled,commentsnumbered]{algorithm2e}
\usepackage{tikz-qtree,tikz-qtree-compat}
\usepackage{tikz}
\usepackage{tabularx}
\usepackage{cite}

\usetikzlibrary{calc,trees,positioning,arrows,fit,shapes,calc}
\tikzset{every tree node/.style={align=center, anchor=north}}
\definecolor{myblue}{RGB}{80,80,160}
\definecolor{mygreen}{RGB}{80,160,80}
\definecolor{LightCyan}{rgb}{0.88,1,1}
\definecolor{antiquewhite}{rgb}{0.98, 0.92, 0.84}
\usepackage{array}
\newcolumntype{P}[1]{>{\centering\arraybackslash}p{#1}}
\newcolumntype{M}[1]{>{\centering\arraybackslash}m{#1}}%
\newcolumntype{C}[1]{>{\centering\let\newline\\\arraybackslash\hspace{0pt}}m{#1}}

\usepackage{subcaption}
\usepackage{url}
\usepackage{array}
\let\chapter\undefined

\usepackage{fourier}
\usepackage{array}
\usepackage{makecell}
\usepackage{ctable}
\usepackage{wrapfig}

\usepackage{changepage}

\usepackage{listings}% http://ctan.org/pkg/listings
\newcolumntype{M}[1]{>{\centering\arraybackslash}m{#1}}

\usepackage{amsthm}

\newcommand{\system}{{\tt JustInTime}}

\newtheorem{theorem}{Theorem}[section]

\newtheorem{example}[theorem]{Example}

\newtheorem{definition}[theorem]{Definition}


\SetAlFnt{\small}

\usepackage[pdfborder={0 0 0}, plainpages, pdfpagelabels=false, pdfstartview=FitH,bookmarks=false]{hyperref}
\usepackage{paralist}

\usepackage{xcolor}
\hypersetup{
	colorlinks,
	linkcolor={red!50!black},
	citecolor={blue!50!black},
	urlcolor={blue!70!black}
}

\newcommand\copyrighttext{%
  \footnotesize \textcopyright 2019 IEEE. Personal use of this material is permitted.
  Permission from IEEE must be obtained for all other uses, in any current or future
  media, including reprinting/republishing this material for advertising or promotional
  purposes, creating new collective works, for resale or redistribution to servers or
  lists, or reuse of any copyrighted component of this work in other works.
  DOI: \href{<http://tex.stackexchange.com>}{10.1109/ICDE.2019.00221}}
\newcommand\copyrightnotice{%
\begin{tikzpicture}[remember picture,overlay]
\node[anchor=south,yshift=10pt] at (current page.south) {\fbox{\parbox{\dimexpr\textwidth-\fboxsep-\fboxrule\relax}{\copyrighttext}}};
\end{tikzpicture}%
}

\begin{document}
\title{Just in Time: Personal Temporal Insights for Altering Model Decisions}

% \author{
%   Naama Boer\\
%   Tel Aviv University\\
%   \texttt{naamaboer@mail.tau.ac.il}
%   \and
%   Daniel Deutch\\
%   Tel Aviv University\\
%   \texttt{danielde@post.tau.ac.il}
%   \and
%   Nave Frost\quad\quad\\
%   Tel Aviv University\quad\quad\\
%   \texttt{navefrost@mail.tau.ac.il}\quad\quad
%   \and
%   Tova Milo\quad\quad\\
%   Tel Aviv University\quad\quad\\
%   \texttt{milo@post.tau.ac.il}\quad\quad
% }
% \date{}

\author{
    \IEEEauthorblockN{Naama Boer}
    \IEEEauthorblockA{Tel Aviv University\\
        naamaboer@mail.tau.ac.il}
    \and
    \IEEEauthorblockN{Daniel Deutch}
    \IEEEauthorblockA{Tel Aviv University\\
        danielde@post.tau.ac.il}
    \and
    \IEEEauthorblockN{Nave Frost}
    \IEEEauthorblockA{Tel Aviv University\\
        navefrost@mail.tau.ac.il}
    \and
    \IEEEauthorblockN{Tova Milo}
    \IEEEauthorblockA{Tel Aviv University\\
        milo@post.tau.ac.il}

}

\maketitle

\copyrightnotice

\begin{abstract}
    The interpretability of complex Machine Learning models is coming to be a critical social concern, as they are increasingly used in human-related decision-making processes such as resume filtering or loan applications. Individuals receiving an undesired classification are likely to call for an explanation - preferably one that specifies what they should do in order to alter that decision when they reapply in the future. Existing work focuses on a single ML model and a single point in time, whereas in practice, both models and data evolve over time:  an explanation for an application rejection in 2018 may be irrelevant in 2019 since in the meantime both the model and the applicant's data can change.

To this end, we propose a novel framework that provides users with insights and plans for changing their classification in particular
future time points. The solution is based on combining state-of-the-art algorithms for (single) model explanations, ones for predicting future models, and database-style querying of the obtained explanations. We propose to demonstrate the usefulness of our solution in the context of loan applications, and interactively engage the audience in computing and viewing suggestions tailored for applicants based on their unique characteristic.
\end{abstract}

\section{Introduction}

\label{intro}

%societal concerns in data management have been receiving increasing amounts of interest and attention. As time goes by,

In recent years critical decision-making processes, such as credit
score assignment, resume filtering and loan approvals
are vastly supported by complex machine learning classifiers.  A
significant drawback to the utilization of such complex models is
their opaqueness, which leaves two important questions unanswered: why did we get a classification result, such as a loan request rejection? What can we do to change it, e.g., what could an applicant do so that his next loan request is approved?

To address this problem, much research has been devoted recently to
the explainability of complex Machine Learning models. For example,
some solutions \cite{datta2016algorithmic,shrikumar2017learning} identify the most influential features in the input, and others \cite{ribeiro2016should} produce approximated interpretable models.
These approaches typically provide answers only as to {\em why a decision was made}.
However, they fall short of providing a {\em practical plan of
action for changing the decision}. This is in part because they do
not account for the temporal aspect of such a plan, namely the
evolution of both the user characteristics and the classification
models.

\begin{example}
    \label{ex:reject}
To illustrate, consider a bank that uses a Machine Learning model to
classify loan application based the following features: \emph{Age, Household status, Annual Income, Monthly Debt, Job
	Seniority} and the requested \emph{Loan Amount}. Further consider an
applicant, John, 29 years old, whose loan request is rejected. 

John wants to come up with a plan for altering this rejection. By employing a tool such as \cite{datta2016algorithmic}, he is advised to increase his income by 20\%. John works hard to get promoted at work and achieves the desired income within two years. He then reapplies, only to discover that the loan criteria have
changed, and his request is again rejected. This time the
explanation is that his debt is too high. 
He wasn't aware that for people over 30, income requirements are often relaxed while debt requirements tend to become stricter. In hindsight, John would rather focus on decreasing his debt over working towards getting a raise.
\end{example}

The suggestions given to the user in this simple example suffer
from several deficiencies. They ignore the expected changes in the
user's profile (e.g. age increases over time, and often so does
seniority), as well as potential changes in the
classifier, due to e.g varying data
distributions, some of which may be predicted based on past
behavior. Furthermore, users differ in their preferences and limitations. For example, one may not be able to increase salary
beyond a certain bound, or would rather decrease debt instead. 
%Such characteristics must be taken into account in order to produce a practical plan. 
In a dynamic user-focused environment, a satisfactory solution must guide the user in three fundamental aspects: Which features to modify, how to modify them, and when to (re)apply. Following are examples of questions related to these aspects, which are likely to be raised by applicants:

\begin{enumerate}
    \item \textbf{No modification:} What is the closest time point (if any) at which reapplying without modifications will be APPROVED?

   	\item \textbf{Minimal features set:} What is the smallest set of features whose modification can lead to APPROVAL? (when? and how should they be modified?).

  	\item \textbf{Dominant feature:} Is there a single feature whose modification leads to APPROVAL in all future time points? (and how should it be modified at each point?)

   	\item \textbf{Minimal overall modification:} What is the minimal overall modification (by some distance measure) that leads to
    APPROVAL, and when?

    \item \textbf{Maximal confidence:} Which modifications (and at which time point) would maximize chances of APPROVAL?

%    \begin{example}
%        An applicant is interested in maximizing the chances of her loan approval. A suitable suggestion would be to reapply again next year, with income increased in at least 3,000\$ and her monthly debt decreased in 400\$ (by refinancing her mortgage for example).
%    \end{example}
%    \begin{example}
%        An applicant is interested in a minimal modification that would facilitate her loan approval. A suitable insight would suggest that without any modifications, within two years (due to changes in the applicant's age and seniority) the application is expected to be approved.
%    \end{example}

    \item \textbf{Turning point:} Is there a time point after which, with some modifications, the confidence of being APPROVED always exceeds $\alpha$?

%    \item \textbf{Maximal acceptance frame:} What is the maximal time frame with potential modifications leading to APPROVAL?

\end{enumerate}

We present \system~, a system enabling users to obtain answers to such questions, thereby assisting them in devising a practical plan of action for altering undesired decisions. The system relies on past labeled data with timestamps.
%(e.g past loan applications are labeled as 'fully paid' or 'charged off'). 
An initial configuration is performed by a system administrator through a dedicated \emph{admin UI}.
The administrator sets parameters controlling the amount and time intervals between future time points. According to these parameters the \emph{Models Generator} processes the training data and trains a
sequence of models, each designated for a specific future time span.
In addition, the administrator may define global \emph{Domain constraints} derived from the domain characteristics (such as database integrity constraints), that will be imposed on all users. 

User interactions use a dedicated UI and begin with defining preferences and limitations in the form of constraints. The \emph{Candidates Generator} components then generate a set of {\em modifications candidates} for each time point, and store them in the system's database. Each of these candidates represents a potential applicable modification to the user's data profile, that is expected to alter
the corresponding model's decision if applied. Users can then query the database through a friendly dedicated interface, consisting of canned questions of the type exemplified earlier (translated to SQL queries)\footnote{Experts users may compose additional SQL queries}, with the output presented to the user in the form of
verbal or graphic insights.

We will demonstrate the operation of \system~ through a {\em loan
application} scenario, using the \emph{Lending Club Loan Data} public dataset containing the details of approximately 1M loan applications from the years 2007-2018 \cite{lendingclub}. Specifically we will
demonstrate how applicants whose applications got rejected can obtain actionable
insights using a variety of queries. The demonstration will
interactively engage the audience who will play the role of rejected applicants, showcasing the system's operation and its different facets.

\begin{figure}[!t]
	\begin{minipage}[b]{\linewidth}\centering
		\includegraphics[width=\linewidth]{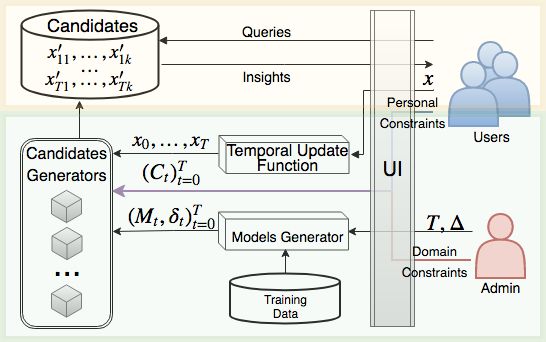}\vfill
	\end{minipage}
	\caption{System Architecture}
	\label{fig:sysarch}
\end{figure}

\section{Technical Overview}
\label{sec:tech}

In this section, we formally define our framework components and
demonstrate their operation. We start by providing the
necessary preliminaries for the case of a single decision-making
model at the present time. Afterwards, we introduce our temporal
framework that consists of two phases. In the first phase, a database of
candidate proposals is generated, and in the second phase the user interacts
with the data gathered in order to obtain insights for
achieving the desired classification.

\subsection{Preliminaries}
We start by recalling some basic notions required for the generation
of proposals. A key notion that we use is that of a
\emph{Machine Learning model}. For simplicity, we focus on binary
classification, but the framework can be easily generalized to
multi-class problems.

\begin{definition}
    Let $d$ be the dimension of the input space, a \emph{Machine Learning model} $M:\mathbb{R}^d \mapsto [0,1]$ is a function such that $\forall x \in \mathbb{R}^d$, $M(x)$ is the probability of the desired positive classification of the vector $x$.
\end{definition}

In our setting, the vector $x$ represents the user's attributes
(i.e. profile). For altering a decision, some of these attributes 
should be modified. However, as observed in
\cite{deutch2019}, in practical settings not all modifications are viable.
For example, a person's age can not be decreased, a user may prefer not to change her address, etc. Thus, users may further
define a constraints function according to their preferences and limitations.

%Recall example \ref{ex:reject} in which an applicant's loan request
%was denied, and the advice he was given did not match his
%preference. In \cite{deutch2018, deutch2019} it was observed that
%not all modifications are viable, thus users are given an
%opportunity to define a constraint function according to their
%personal preferences and limitations.

\begin{definition}
    A \emph{Constraints Function} $C$ maps vectors in $\mathbb{R}^d$ to a subset of $\mathbb{R}^d$. For a given input $x \in \mathbb{R}^d$, the set $C(x) \subseteq \mathbb{R}^d$ denotes the valid modifications to $x$.
\end{definition}

%\textcolor{orange}{
%\emph{System Constraints} are defined by the administrator, who uses domain knowledge to formalize global characteristics of the model input and its behavior over time. Users may define any amount of \emph{User Constraints} according to their personal preferences and limitations. These two types of constraints are joined to form the \emph{Constraint Function} defined above.}

In \system{}, constraints specified by the administrator and the
user are joined to form a general constraints function as defined
above. Constraints may refer to a single point in time or all of
them, and may contain any number of linear inequalities
joined by conjunctions and disjunctions, over any subset of 
attributes of the input vector. In addition to user attributes, constraints can refer to three special properties: distance from the input w.r.t. $l_2$-norm (termed 'diff'), distance w.r.t. $l_0$-norm ('gap'), and the overall model score ('confidence') of the given input.

\begin{definition}
\label{ac}
    Given a model $M$, input $x$ and constraints $C$ the set of \emph{Decision Altering Candidates} is defined as:
    $$A = \{x' \in \mathbb{R}^d \; \mid \; x' \in C(x) \; and \; M(x') > \delta\}.$$
    Where $\delta$ is the model threshold, i.e. candidates with model score greater then $\delta$ are classified positively.
\end{definition}

We base our solution on an algorithm developed in \cite{deutch2019}
by a subset of the present authors, for finding a candidate $x' \in A$ that
aims to minimize the distance from the original input point $x$.
Obtaining an optimal candidate was shown to be an NP-hard problem
for e.g Random Forests and Neural Networks, hence the iterative
algorithm developed in \cite{deutch2019} applies model-dependent
heuristics. While convergence is not guaranteed in general, it was shown empirically that the algorithm converges after a small number of iterations. We adjusted the algorithm to our problem by incorporating diverse objectives (confidence, gap and diff) when searching for the candidates, as opposed to a single distance measure. In addition, we output top-$k$ candidates in each iteration, as opposed to just one, using a beam search with width $k$ to prune the least promising candidates. 

For reasons explained in the introduction, considering only the
modification that would alter the decision in the {\em present}
model is insufficient, thus in the following subsection we suggest
a generalized framework that takes the temporal aspect into
consideration.

\setlength{\textfloatsep}{0pt}

\begin{figure}[!htb]
	\scriptsize
	
	\begin{minipage}{1.0\columnwidth}
		
		\begin{tabular}{| c | c |}
			\specialrule{.15em}{0em}{0em}
			\makecell{\textbf{\textsuperscript{(1)}No} \\ \textbf{modification}}
			&
			\begin{tabular}{l}
				\verb|SELECT Min(time)                           |\\
				\verb|FROM candidates|\\
				\verb|WHERE diff = 0|\\
			\end{tabular}\\
		
			\specialrule{.05em}{0em}{0em}
			\makecell{\textbf{\textsuperscript{(2)}Minimal} \\ \textbf{features set}}
			&
			\begin{tabular}{l}
				\verb|SELECT *                                   |\\
				\verb|FROM candidates|\\
				\verb|ORDER BY gap|\\
				\verb|LIMIT 1|\\
			\end{tabular}\\

			\specialrule{.05em}{0em}{0em}
			\makecell{\textbf{\textsuperscript{(3)}Dominant} \\ \textbf{feature}\\
				\textbf{(income)}}
			& \begin{tabular}{l}
				\verb|SELECT distinct time as t|\\
				\verb|FROM candidates|\\
				\verb|WHERE EXISTS|\\
				\verb|      (SELECT * |\\
				\verb|       FROM candidates as cnd|\\
				\verb|       INNER JOIN temporal_inputs as ti|\\
				\verb|               ON ti.time = cnd.time|\\
				\verb|       WHERE cnd.time = t|\\
				\verb|             AND ((gap = 0) OR (gap = 1 |\\
				\verb|             AND cnd.income != ti.income)))|\\
			\end{tabular}
			\\

			\specialrule{.05em}{0em}{0em}
			\makecell{\textbf{\textsuperscript{(4)}Minimal} \\
				\textbf{overall} \\
				\textbf{modifications}}
			&
			\begin{tabular}{l}
				\verb|SELECT Min(diff)                           |\\
				\verb|FROM candidates|\\
			\end{tabular}\\		
			
			\specialrule{.05em}{0em}{0em}
			\makecell{\textbf{\textsuperscript{(5)}Maximal} \\ \textbf{confidence}}
			&
			\begin{tabular}{l}
				\verb|SELECT *                                   |\\
				\verb|FROM candidates|\\
				\verb|ORDER BY p DESC|\\
				\verb|LIMIT 1|\\
			\end{tabular}\\
			
%			\specialrule{.05em}{0em}{0em}
%			\makecell{\textbf{\textsuperscript{(6)}Minimize} \\ \textbf{Features} \\ \textbf{amount}}
%			&
%			\begin{tabular}{l}
%				\verb|SELECT Min(gap)                            |\\
%				\verb|FROM candidates|\\
%			\end{tabular}\\										
			
			\specialrule{.05em}{0em}{0em}
			\makecell{\textbf{\textsuperscript{(6)}Turning} \\ \textbf{time point}}
			& \begin{tabular}{l}
				\verb|SELECT Min(time)                           |\\
				\verb|FROM candidates|\\
				\verb|WHERE time >= ALL|\\
				\verb|      (SELECT time as t|\\
				\verb|       FROM candidates|\\
				\verb|       WHERE EXISTS (...))|\\
			\end{tabular}
			\\
%			\specialrule{.05em}{0em}{0em}
%			\makecell{\textbf{\textsuperscript{(7)}Maximize} \\ \textbf{acceptance} \\ \textbf{frame}}
%			& \begin{tabular}{l}
%				\verb|SELECT c1.time, c2.time as t1, t2          |\\
%				\verb|FROM candidates as c1, candidates as c2|\\
%				\verb|WHERE t1 <= t2 AND NOT EXISTS (|\\
%				\verb|      SELECT time|\\
%				\verb|      FROM candidates|\\
%				\verb|      WHERE time between t1 AND t2|\\
%				\verb|            AND NOT EXISTS (|\\
%				\verb|                            ...)|\\
%			\end{tabular}
%			\\
			\specialrule{.15em}{0em}{0em}
			
		\end{tabular}
	\end{minipage}
	
	\caption{Sample of predefined queries}
	\label{fig:queries}
\end{figure}

\setlength{\textfloatsep}{0pt}

\begin{figure*}[ht]
	\centering
	\begin{subfigure}{0.45\textwidth}
		\includegraphics[width=\textwidth]{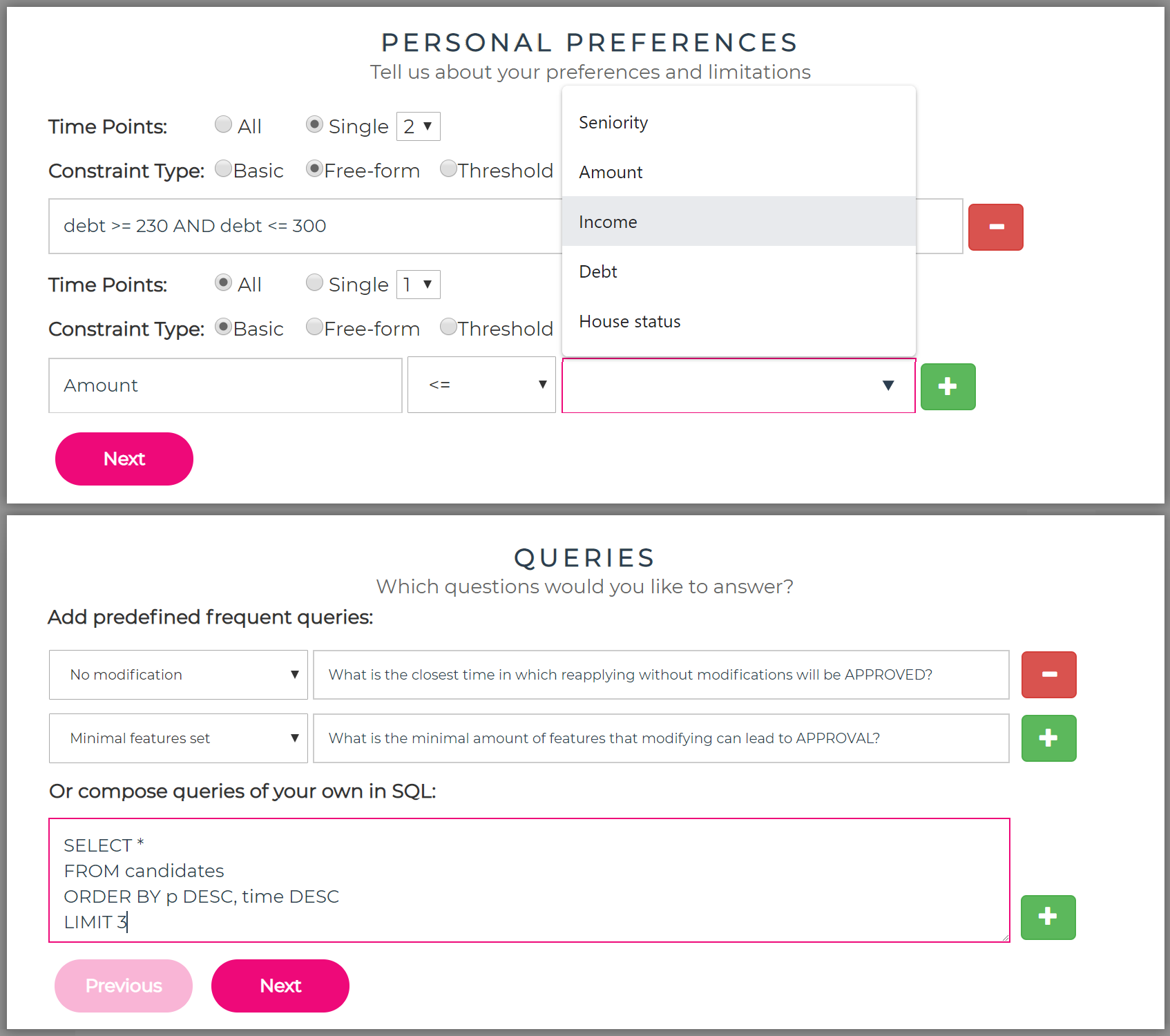}
		\caption{Preferences and Queries Screens}
		\label{fig:constraints_queries}
	\end{subfigure}
	\begin{subfigure}{0.45\textwidth}
		\includegraphics[width=\textwidth]{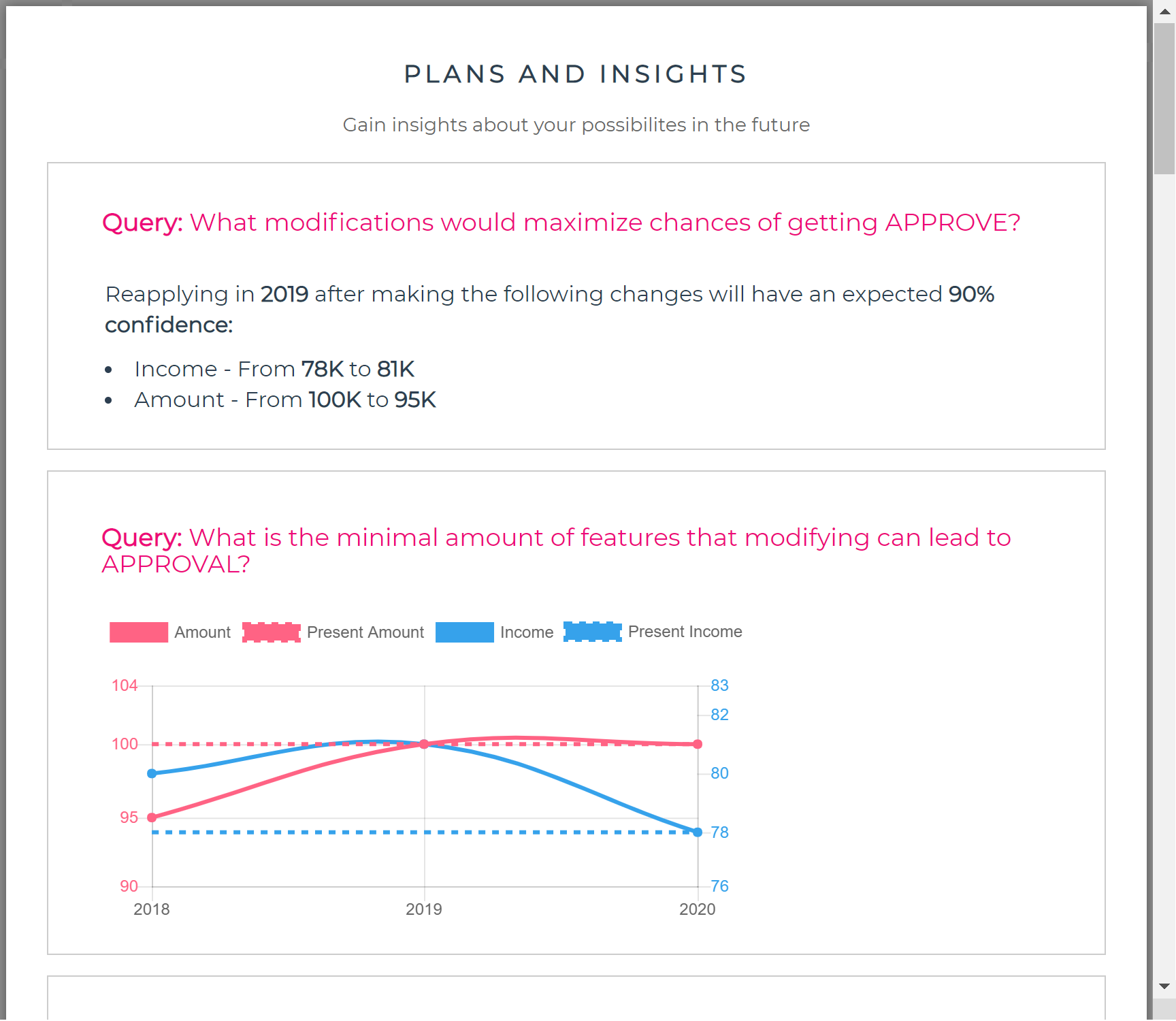}
		\caption{Insights Screen}
		\label{fig:insights_screen}
	\end{subfigure}
	\caption{User Interface Screen-shots}
	\label{UIScreens}
\end{figure*}

\subsection{Temporal Candidates Generation}
The temporal candidates generation phase is  shown in the bottom
section of Figure \ref{fig:sysarch}. In order to provide users
with helpful insights regarding the future, the system has to predict how
future models are likely to operate.
%Hence, the first step in this process is generation of \emph{Machine Learning models}, each relevant to a certain future period.
Generating future models requires labeled training data with
timestamps. Additionally, two parameters, $T$ and $\Delta$, determine the
time span handled by the system. $T$ is the number of time points
considered and $\Delta$ sets the length of the interval between consecutive
points. The models generator then uses existing domain adaptation methods \cite{lampert2015predicting}, in order to
create a sequence of pairs $(M_t, \delta_t)_{t=0}^T$, where $M_t$ is
the expected approximated model at future time $t$, and $\delta_t$
is its threshold. The models generator uses the training data to
learn the time variations of the data distribution, using a method
from \cite{lampert2015predicting}, that relies on two techniques:
probability distribution embedding into a reproducing kernel
Hilbert space, and vector-valued regression. Future models are
then trained based on the approximated future distributions. Note that this part of the candidates generation process is performed once and is independent of any specific user.

Recall that some of the attributes (features) are known to change over time (e.g. age increases). Thus, every
future model operates on future representations of the features vector, which is generated in our setting using a
{\em Temporal Update Function}.

\begin{definition}
    A \emph{Temporal Update Function} receives a vector $x \in \mathbb{R}^d$. For features specified as \emph{``non temporal``} $f$ is the identity function. For every \emph{``temporal"} feature $v$, the value of $v$ at time point $t$ is given by  $f(x,t)[v]$.
\end{definition}

\begin{example}
To continue with our running example, $age$ is a temporal feature hence $f(x,3)[age] = x[age] +
3\Delta$.
%On the other hand, $birth\_date$ is non temporal hence $f(x,t)[birth\_date] =
%x[birth\_date]$ for every $t$.
%   See table \ref{tab:temporal_updated_data}.
\end{example}

As previously mentioned, each user is represented by a vector
$x\in\mathbb{R}^d$. The temporal update function operates on
$x$ and outputs the future temporal representations of $x$, namely
$\{f(x,t)\}_{t=0}^T$, denoted $x_0, \dots, x_T$, for simplicity. These
outputs are stored in a relational table called \emph{temporal
inputs}. 
%\tova{is it important that it is in a relational table?
%isn't it enough to say that we refer to them by this name.} 
The input to the candidates generators is then:

\begin{itemize}
    \item Temporal input vectors $x_0, \dots, x_T$ in $\mathbb{R}^d$.

    \item Sequence of model and threshold pairs $(M_t, \delta_t)_{t=0}^T$.

    \item Constraints functions $(C_t)_{t=o}^T$ which are the conjunction of user and domain
    constraints. 
\end{itemize}

The core of the temporal candidates generation phase is a
sequence of candidates generator components presented in the
previous subsection. Each generator is responsible for outputting $k$
decision-altering candidates fitted for the corresponding time point
$t \in \{0, \dots, T\}$, that we denote $x'_{t1}, \dots, x'_{tk}$. The
generators are independent of each other, and thus they can be executed in parallel. The results (set of
candidates per time point) are then stored in a \emph{candidates} table,
ready to be queried by the user.

Since $A_t$, the set of decision altering candidates (Definition
\ref{ac}) at time $t$, may be arbitrarily large, whereas we are interested in a small, optimized and diverse subset per each time point, we
employ the aforementioned adaptation of the algorithm in \cite{deutch2019} to select a diverse set of top-$k$ candidates, using several useful predefined metrics. The diversity ensures that
limiting the number of candidates does not lead to a
degradation in the quality of the answers to user queries.

%\begin{example} Table \ref{tab:model_min_changes} presents the top-3 decision-altering candidates that the applicant needs to take every year, in order to get the loan approved. Note that the diff listed in the table is calculated after scaling the data. The maximal confidence and minimal gap and diff modifications are highlighted for every time-point.
%\end{example}

\subsection{Obtaining insights for Altering Model Decisions}

The previous phase is concluded with the generation of the \emph{candidates} table. In this phase the user
interacts with the relational database in order to obtain
personalized insights. Experts may interact with the system directly in SQL, while for non-experts we provide an expressive and easy-to-use interface for formulating a broad range of common questions (including those listed in the introduction), translated by the system into SQL queries. Figure \ref{fig:queries} exemplifies several interesting
questions along with their corresponding SQL queries.

\section{System and Demonstration Overview}
\label{sec:demo}

The \system{} backend is implemented in Python 3.6 and its frontend in JavaScript. It uses MySQL server for storing and querying the generated database.
The candidates generator and the models generator components use the \emph{H2O} module for data modeling. The models generator trains a random forest classifier for each time span.

The operation of the system is demonstrated w.r.t. the \emph{Lending Club Loan Data} containing details of approximately
$1M$ loan applications from 2007 to 2018, published in Kaggle \cite{lendingclub}. 
%In this example use case, the \emph{Models Generator} uses $H2O$ to train \emph{Random Forest} classifiers with 20 trees and depth of 6. 
We will demonstrate that \system{} helps loan applicants to obtain actionable insights on a variety of questions. The demonstration will interactively engage the audience, showcasing the different facets of the system. 

%In the first step, users are asked to define preferences and limitations. Each preference in the \emph{Personal Preferences} screen (depicted in figure \ref{fig:constraints_queries}) may be one of the following:

We will start by demonstrating the functionality of the system through a reenactment of five real-life loan applications that were denied. The interaction with the system is performed in three steps, each in a designated screen. 
%Users interaction with the system starts in the \emph{Peronal Prefrences} screen (depicted in Figure \ref{fig:constraints_queries}) where they are asked to define preferences and limitations in the form of constraints. 
The audience will suggest preferences and limitations on behalf of the rejected applicants in the \emph{Personal Preferences} screen. In the \emph{Queries} screen, we offer the audience a selection of predefined queries, such as the ones listed in the introduction (bottom of Figure \ref{fig:constraints_queries}). The \emph{Plans and Insights} screen presents insights with regard to these queries, subject to conditions set in the preferences step (Figure \ref{fig:insights_screen}).

We will continue the demonstration prompting participants to play the role of loan applicants. After setting their features, they will be assigned a score by the present classifier, and further interact with the system in order to attain a plan that achieves their goals. Since the system was designed for non-experts, it requires no prior skills.

To illustrate what happens behind the scenes, we will 
%also 
invite the audience to observe the operation of different stages in our solution. Starting with an excerpt of the raw training data, and continuing with temporal representations of input vectors and generated future models. 
Lastly, we will examine the execution of a single candidates generator and the candidates it generates.

\textbf{Related work} As the main focus of our work is to provide explanations within a temporal framework, it relates to both  \emph{Interpretable Machine Learning} and \emph{Domain Adaptation}. A significant part of the work on complex models explainability focuses on explaining particular predictions, often in the context of social accountability and transparency. Within this class of works, the closest to ours are such that use perturbations of the test point and draw conclusions based on how the prediction changes \cite{datta2016algorithmic,adler2018auditing,deutch2019}. These solutions mostly promote understanding of a specific decision and engender trust in the model performance, but they are not well suited for attaining an actionable plan for altering the classification, subject to personal and temporal constraints. 

Our work also closely relates to the field of Domain Adaptation, specifically the task of predicting future classifiers given past labeled data with timestamps \cite{kumagai2016learning, lampert2015predicting}. These are complementary to our work, as we use such %domain adaptation 
solutions as a component of our framework.

\begin{small}
	\section*{Acknowledgment}
	This work has been partially funded by the Blavatnik Fund, the Israeli Science Foundation, Intel, and the European Research Council (ERC) under the Europe Unions Horizon 2020 research and innovation programme (grant agreement No. 804302).
\end{small}

\begin{small}
    \bibliographystyle{IEEEtran}
    \bibliography{IEEEabrv,bibShort}
\end{small}

% \newpage
%\appendix
% \input{letter}

\end{document}